# Transition from Mott insulator to superconductor in GaNb$_4$Se$_8$ and GaTa$_4$Se$_8$ under high pressure

M. M. Abd-Elmeguid[1], B. Ni[1], D. I. Khomskii[1,a], R. Pocha[2], D. Johrendt[2], X. Wang[3], and K. Syassen[3]

[1]II. Physikalisches Institut, Universität zu Köln, Zülpicher Str. 77, 50937 Köln, Germany
[2]Department Chemie, Ludwig-Maximilians-Universität München, Butenandtstr. 5-13 (Haus D), 81377 München, Germany
[3]Max-Planck-Institut für Festkörperforschung, Heisenbergstr. 1, 70569 Stuttgart, Germany

Electronic conduction in Ga$M_4$Se$_8$ ($M$ = Nb, Ta) compounds with the *fcc* GaMo$_4$S$_8$-type structure originates from hopping of localized unpaired electrons ($S = \frac{1}{2}$) among widely separated tetrahedral $M_4$ metal clusters. We show that under pressure these systems transform from Mott insulators to a metallic and superconducting state with $T_C$ = 2.9 K and 5.8 K at 13 GPa and 11.5 GPa for GaNb$_4$Se$_8$ and GaTa$_4$Se$_8$, respectively. The occurrence of superconductivity is shown to be connected with a pressure-induced decrease of the $M$Se$_6$ octahedral distortion and simultaneous softening of the phonon associated with $M-$Se bonds.

Superconductivity in the presence of strong electron correlations has attracted considerable attention, especially after the discovery of high-$T_C$ superconductors. Usually superconductivity is obtained in such systems by doping Mott insulators, like in cuprates [1] or in $Na_xCoO_2 \cdot yH_2O$ [2]. Another option is to study the occurrence of superconductivity under high pressure in *stoichiometric* systems in the proximity to a Mott transition. The advantage in this case is the absence of disorder. Unfortunately, there are very few such systems known (e.g. $\beta$-$Na_{0.33}V_2O_5$ [3] and recent theoretical discussion [4]).

In this work we show that cluster compounds $GaM_4Se_8$ ($M$ = Nb, Ta), which are nonmagnetic Mott insulators at ambient pressure, transform to a metallic and superconducting state at pressures of 13 GPa and 11.5 GPa with critical temperatures $T_C$ = 2.9 K and 5.8 K, respectively. We show that the Mott transition itself is apparently connected with internal distortions of the clusters rather than a change of the lattice symmetry. We also observed a rather strong softening of one of the phonon modes, which correlates with the appearance of superconductivity.

Ternary chalcogenides $AM_4X_8$ ($A$ = Ga, Ge; $M$ = V, Mo, Nb, Ta; $X$ = S, Se) belong to an interesting class of transition metal systems which exhibit strong electronic correlation effects. The origin of the electronic correlation in these systems is a consequence of their peculiar crystal structure, shown in Fig. 1(a). This *fcc* structure ($GaMo_4S_8$-type) can be described as a deficient spinel $A_{0.5}M_2X_4$ [5,6], in which the ordering of the tetrahedral $A$-ions reduces the symmetry from $Fd\bar{3}m$ to $F\bar{4}3m$. As a result, the $M$ (transition metal)-atoms are shifted off the centers of the S / Se octahedral, see Fig. 1(b), forming tetrahedral $M_4$-clusters with typical intracluster $M-M$ distances $(d_M)$ of ≤ 3 Å. At the same time the $M-M$ distances $(d_C)$ between the $M_4$-clusters become large (> 4 Å), which results in a formation of localized electronic states in the clusters. This leads to unusual transport and magnetic properties. None of these compounds show metallic conductivity; instead the electronic conduction takes place by hopping of carriers between the clusters [7-10]. Simultaneously, magnetic susceptibility is typical for localized spins. Thus, this class of systems can be considered as Mott-insulators.

The ground state properties of these compounds strongly depend on the *local* electronic structure of the $M_4$-cluster (actually $M_4X_4$-clusters) which is mainly determined by the number of valence electrons per cluster [9-11]. According to molecular orbital (MO) calculations, the *d*-orbitals within the $M_4$-clusters can be described by MO's which consist of three energetically different bonding states (for cubic $T_d$ symmetry) [12]: a nondegenerate level ($a_1$), followed by twofold ($e$) and threefold ($t_2$) degenerated levels (see Fig. 1(c)). For cluster compounds of the type $Ga^{3+}(M^{3.25+})_4(S^{2-},Se^{2-})_8$, we have 7 valence electrons per cluster with $M$ = V, Nb, Ta and 11 electrons with $M$ = Mo. In both cases, the occupation of the cluster orbitals leads to *one unpaired* electron (i.e. $S$ = ½) per cluster. This is in agreement with the values of the magnetic moments obtained from magnetic susceptibility measurements and is also consistent with spin polarized band structure calculations [10,11]. In one respect, however, these systems are different from the conventional Mott insulators such as transition metal oxides: in contrast to the latter, the correlated units are $M_4$ clusters which may have extra internal degrees of freedom. As we will show below, this leads to a high sensitivity of these systems to external pressure.

Single phase polycrystalline samples of $GaTa_4Se_8$ and $GaNb_4Se_8$ were prepared as described in Ref. [9]. X-ray powder patterns were completely indexed using the structural data obtained from single crystal experiments [13]. The pressure dependence of the lattice constants at 300 K up to about 26 GPa was measured on powdered samples by energy dispersive x-ray diffraction (EDX) at HASYLAB using the diamond anvil cell (DAC) technique. The same type of DAC has been used for conventional four-terminal electrical resistance measurements up to about 29 GPa between 1.6 K and 300 K. Single crystal x-ray diffraction measurements (MoK$_{\alpha 1}$) of $GaTa_4Se_8$ were performed at 300 K up to $p$ = 15 GPa using a special DAC. Raman spectra were recorded in back-scattering geometry using a micro-spectrometer.

Before discussing the high pressure results we briefly mention some experimental data at ambient pressure. The values of the lattice parameter $a$ as determined from x-ray diffraction measurements at 300 K are found to be 10.440(1) Å and 10.358(1) Å for $GaNb_4Se_8$ and $GaTa_4Se_8$, respectively, in agreement with previous results [6]. From single crystal x-ray data we obtained the values of the characteristic intra- and intercluster distances: $d_M$ = 3.051(3) Å, 3.015(2) Å and $d_C$ = 4.332(3) Å, 4.338(2) Å, for $GaNb_4Se_8$ and $GaTa_4Se_8$, respectively. Measurements of the temperature dependence of electrical resistivity (1.6 K ≤ $T$ ≤ 300 K) show for both samples a semiconductor-like behavior with activation energies of 0.14 eV ($GaNb_4Se_8$) and 0.1 eV ($GaTa_4Se_8$). Actually, the activation energy decreases with decreasing temperature, in agreement with that reported for $GaMo_4S_8$ and $GaV_4S_8$ [10]. The magnetic susceptibility of the two samples shows Curie-Weiss behavior (100 K ≤ $T$ ≤ 300 K), indicating the existence of magnetic correlations, but no magnetic ordering is found down to 1.6 K in agreement with Ref. [6]. The estimated values of the effective magnetic moments are 1.6 $\mu_B$ per Nb$_4$-cluster (close to theoretical value 1.73 $\mu_B$ for $S$ = ½) and 0.7 $\mu_B$ per Ta$_4$- cluster. Detailed analysis of the results at ambient pressure are presented elsewhere [13,14], in the present paper we focus on high pressure results.

Fig. 2(a, b) displays the temperature dependence of the normalized electrical resistance $R_n = R(T)/R(297K)$ in the temperature range 1.6 K $\leq T \leq$ 300 K as a function of pressure for GaNb$_4$Se$_8$ and GaTa$_4$Se$_8$, respectively. Considering first the overall behavior of $R_n(T,p)$ in both samples, one finds with increasing pressure a gradual change from the semiconducting to a metallic-like behavior and a sudden drop of $R_n$ at low temperatures above a critical pressure ($p_c$), indicative of a superconducting transition. While the metallic behavior $dR/dT > 0$ is observed at rather high pressures ($p \geq$ 19 GPa (GaNb$_4$Se$_8$) and $p \geq$ 15 GPa (GaTa$_4$Se$_8$)), superconductivity already sets in at lower pressures where the temperature dependence of $R_n$ is still semiconducting-like; $T_C$ = 2.9 K at 13 GPa for GaNb$_4$Se$_8$ and 5.8 K at 11.5 GPa for GaTa$_4$Se$_8$. This type of behavior is usually observed in the superconducting state of polycrystalline sintered samples, e.g. at ambient pressure in La$_{1-x}$Sr$_x$CuO$_4$ [15] and under high pressure in the Chevrel phase compound Eu$_{1.2}$Mo$_6$(S,Se)$_8$ [16, 17], and is known to be due to a coexistence of superconducting and semiconducting phases (granular superconductivity), in the bulk and surface of the grains of such samples, respectively. This explains the finite value of the resistivity observed in the superconducting state of our samples ($\rho_0 \approx 10^{-4} \Omega cm$ at $T$ = 1.6 K and $p \approx$ 20 GPa) despite their single phase purity as well as the increase of the drop of $R(T)$ with increasing pressure (see Fig. 2). We note, however, that the drop of $R(T)$ is substantial (~ 70 %) at $p \approx$ 20 GPa and 1.6 K and is expected to further increase at lower temperatures resulting in a lower value of the resistivity. This indicates an increase of the fraction of superconductivity in the samples with increasing pressure. Fig. 3(c) shows the pressure dependence of $T_C$ for both samples. The value of $T_C$ increases remarkably with increasing pressure up to $p \approx$ 22 GPa, $\partial T_C/\partial p \approx$ 0.4 KGPa$^{-1}$ and $\approx$ 0.2 KGPa$^{-1}$, for GaNb$_4$Se$_8$ and GaTa$_4$Se$_8$, respectively. For GaNb$_4$Se$_8$ we observe a decrease of $T_C$ with increasing pressure at $p >$ 22 GPa. As we show below this decrease is probably connected with a pressure-induced structural distortion in GaNb$_4$Se$_8$.

To prove that the observed behavior is indeed a superconducting transition, we investigated the effect of external magnetic fields $(B_{ex})$ on the temperature dependence of the electrical resistance at pressures of $p$ = 20 GPa (GaNb$_4$Se$_8$) and $p$ = 22 GPa (GaTa$_4$Se$_8$). Fig. 3(a, b) shows the temperature dependence of the electrical resistance as a function of $B_{ex}$. As expected, $T_C$ is clearly shifted to lower temperatures with increasing $B_{ex}$; the overall behavior of $R(T, B_{ex})$ is indeed typical for a bulk superconducting transition: we find a linear decrease of $T_C$ with $B_{ex}$ that can be described by the well known WHH theory for "dirty superconductors" [18]. This excludes the existence of weak links and/or a minority phase superconductivity. We obtain for GaNb$_4$Se$_8$ and GaTa$_4$Se$_8$ values of the upper critical field $B_{C2}$ ($T \to 0$) of 1.7 T at 20 GPa and 8 T at 17 GPa and a corresponding coherence length $\zeta$ of 130 Å and 61 Å, respectively. Thus, despite the fact that without measurements of the Meissner effect under high pressure, we cannot give a value for the superconducting fraction in our samples, the above mentioned experimental results clearly verify a pressure-induced transition from a Mott-insulating state to a metallic and superconducting state.

Next, we investigate whether the observed pressure-induced superconductivity is connected with a structural instability and/or a change of the lattice dynamics under high pressure. The pressure dependence of the lattice parameter $a$ of the two samples is displayed in Fig. 4(a). As evident from the figure, there is no indication of a structural phase transition up to ~ 20 GPa within the resolution of the EDX measurements. Thus, the onset of superconductivity in both systems is not connected with a structural transition [19]. From the analysis of the obtained data (Fig. 4(a)), we find that the intercluster distance $d_C$ decreases by only about 0.20 Å at $p \approx 20$ GPa (s. Fig. 4(b)), resulting in values of $d_C \approx 4.10$ Å for both compounds. This is still much larger than the Nb-Nb or Ta-Ta intracluster distances ($\approx 3$ Å) and most likely too long for any noticeable overlap of d-orbitals of metal atoms of neighbouring cluster units. Thus, the change of direct metal-metal intercluster hopping with increasing pressure cannot account for the observed metallic and superconducting state.

To obtain specific information about a possible change of the *local structure* with increasing pressure, we have performed single crystal x-ray structure determinations of GaTa$_4$Se$_8$ up to $p$ = 15 GPa. Fig. 4(c) shows the pressure-induced change of the two characteristic Ta-Se bond lengths: Ta–Se1 within the Ta$_4$Se$_4$ clusters and the bridging Ta–Se2 bonds between the clusters (see Fig. 1(b)). As clearly shown in Fig. 4(c), the intercluster bond length Ta–Se2 decreases with increasing pressure more than 3 times stronger than that of the intracluster Ta–Se1. Consequently, the Ta atoms move towards the centre of the TaSe$_6$ octahedra with increasing pressure, which results in a corresponding decrease of the distortion of these octahedral ((see Fig. 1(b)). Apparently, this leads to a strong increase of the hybridization of the 5$d$-states of Ta with the $p$-states of the *bridging* Se2-ions, and to a consequent increase of the effective intercluster hopping. We believe that such a change leads to the observed pressure-induced metallic and superconducting state in $GaNb_4Se_8$ and $GaTa_4Se_8$.

The pronounced changes of the local structure under pressure are expected to affect the vibrational properties, with a possible relationship to the occurrence of superconductivity. Indeed, we find highly unusual changes in phonon frequencies. Selected Raman spectra of $GaTa_4Se_8$ measured up to 15 GPa at 300 K are shown in Fig. 5(a). According to group theory, 12 zone-center Raman-active modes ($3A_1 + 3E + 6T_2$) are expected, but fewer modes are observed. The frequencies of the two well-resolved features seen at ambient pressure are 236 $cm^{-1}$ and 272 $cm^{-1}$. Additional Raman lines become clearly observable with increasing pressure. The most notable effect in the present context is the pressure-driven softening of the strongest Raman band (Fig. 5(b)). The initial shift of -7.3 $cm^{-1}$ saturates near 15 GPa, where the mode frequency has dropped by 20 %. $GaNb_4Se_8$ exhibits qualitatively similar pressure effects; the frequency of the dominant mode decreases in a nonlinear fashion from 234 $cm^{-1}$ at ambient pressure to 180 $cm^{-1}$ at 20 GPa. The soft mode is attributed to vibrations involving the stretching of Ta(Nb)-Se bonds. Polarized single-crystal Raman spectra are needed for a detailed mode assignment.

The pressure-driven mode softening indicates the emergence of a strongly anharmonic potential energy for atomic displacements. The anharmonicity is induced here by the continuous suppression of the octahedral distortion under pressure. Similar, but less pronounced pressure effects on phonon modes have been reported for ionic compounds with distorted octahedral coordination (see e.g. [20]). In view of the changes in chemical bonding discussed above, one may speculate that this soft mode enhances the electron-phonon coupling and may contribute to Cooper pairing.

In conclusion, we observed in cluster compounds $GaNb_4Se_8$ and $GaTa_4Se_8$ a rather rare phenomenon - pressure-induced transition from Mott insulator to a superconductor. We have shown that this transition is not connected with a structural phase transition, but is accompanied, and may be is driven by the *internal* structural modifications - a reduction of the distortion of $MSe_6$ octahedra. The transition to a metallic and superconducting state under pressure is accompanied by strong softening of one of the phonon modes. However, the existing data are not yet sufficient to determine the exact type and mechanism of superconductivity; this requires further study. But the results already obtained make this system a very interesting object for the experimental investigation of the interplay between strong electron correlations and superconductivity in stoichiometric materials without disorder as well as for testing related theoretical models.

M. M. A. would like to thank L. H. Tjeng and K. Westerholt for fruitful discussions. This work was supported by the Deutsche Forschungsgemeinschaft through SFB 608 and JO257/2.

[a] Also at Groningen University, Nijenborgh 4, 9722 AG Groningen, The Netherlands

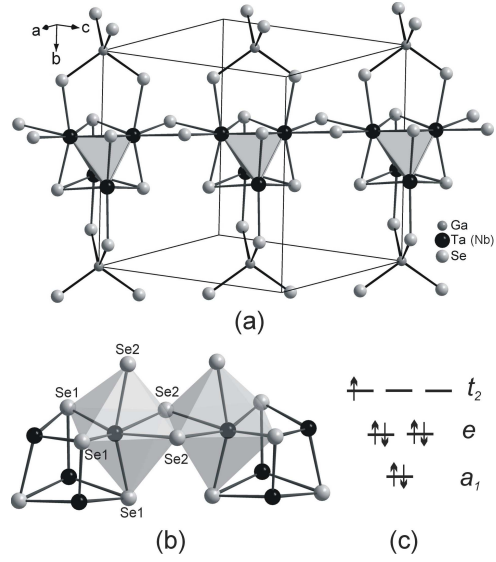

Fig. 1: (a) Linkage of the Ta$_4$Se$_4$ cluster units via bridging Se2 atoms and their connection with the GaSe$_4$ tetrahedra in the *fcc* GaMo$_4$S$_8$ structure. (b) (Ta,Nb) atoms shifted off the centres of distorted edge-sharing Se$_6$ octahedra ($d_{\text{Ta-Se1}} = 2.508$ Å; $d_{\text{Ta–Se2}} = 2.643$ Å). (c) MO-scheme for the $M-M$ bonding orbitals of a $M_4$ cluster with ideal $T_d$ symmetry for 7 electrons per cluster.

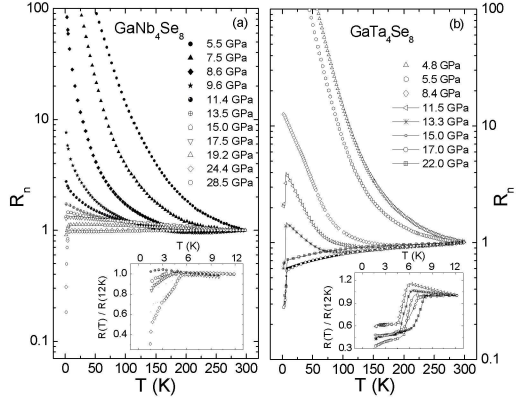

Fig. 2: Temperature dependence of the normalized electrical resistance $R_n = [R(T)/R(297K)]$ of GaNb$_4$Se$_8$ (a) and GaTa$_4$Se$_8$ (b) at different pressures up to 28.5 GPa. Insets show the drop of $R_n$ at high pressures and low temperatures.

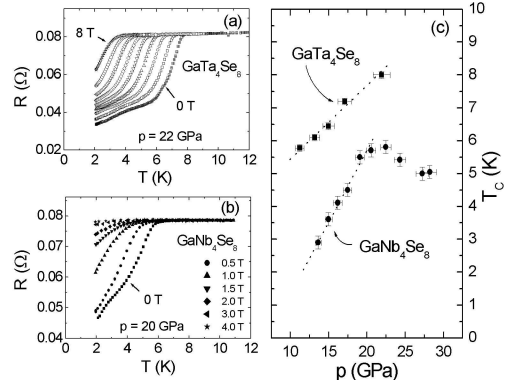

Fig. 3: Temperature dependence of the electrical resistance as a function of magnetic field at pressures of 22 GPa and 20 GPa for GaTa$_4$Se$_8$ (a) and GaNb$_4$Se$_8$ (b), respectively. (c) Pressure dependence of $T_C$ in GaTa$_4$Se$_8$ and GaNb$_4$Se$_8$.

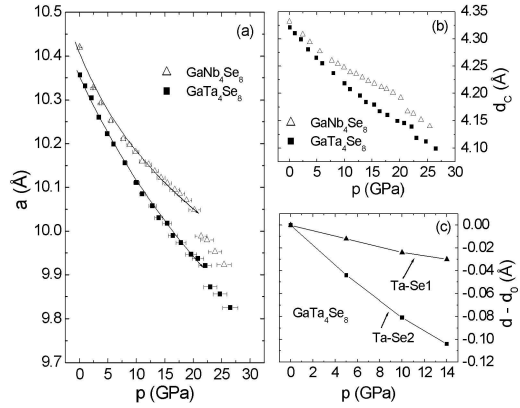

Fig. 4: (a) Pressure dependence of the lattice parameter $a$ of the unit cell of Ga(Nb,Ta)$_4$Se$_8$ at 300 K. (b) Pressure variation of the intercluster distance ($d_C$) of Ga(Nb,Ta)$_4$Se$_8$ as deduced from the experimental data in (a). (c) Pressure-induced change of Ta-Se1 and Ta-Se2 bond lengths in GaTa$_4$Se$_8$ relative to their values at ambient pressure $(d_0)$.

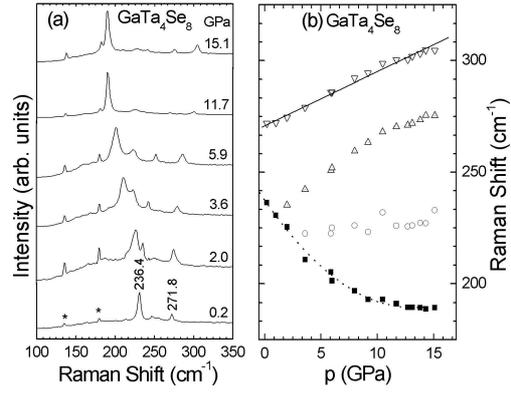

Fig. 5: Raman scattering results for GaTa$_4$Se$_8$: (a) Selected Raman spectra and (b) frequencies of Raman features as a function of pressure. Asterisks in (a) indicate laser plasma lines. Lines in (b) are guides to the eye.